\documentclass[]{iopart}
\usepackage{bm}
\usepackage[latin1] {inputenc}
\usepackage{graphicx}
\newcommand{\be}{\begin{equation}}
\newcommand{\ee}{\end{equation}}
\newcommand{\ba}{\begin{eqnarray}}
\newcommand{\ea}{\end{eqnarray}}

\newcommand{\nn}{\nonumber\\}
\newcommand{\Ref}[1]{(\ref{#1})}

\newcommand{\half}{\textstyle{\frac{1}{2}}}

\begin{document}
\title{The nonlinear fragmentation equation}
\author{Matthieu H. Ernst}
\address{\it Departament de F\'{\i}sica Fonamental, Universitat de
Barcelona, Carrer Mart\'{\i} i Franqu\'es 1, E-08028 Barcelona,
Spain}
\address{Institute for Theoretical Physics,
Universiteit  Utrecht, 3508 TD Utrecht, The Netherlands}
\author{Ignacio Pagonabarraga}
\address{\it Departament de F\'{\i}sica Fonamental, Universitat de
Barcelona, Carrer Mart\'{\i} i Franqu\'es 1, E-08028 Barcelona,
Spain}
\ead{ipagonabarraga@ub.edu}

\begin{abstract}
We study the kinetics of  nonlinear irreversible fragmentation.   Here fragmentation is induced by
interactions/collisions between pairs of particles, and modelled by
general classes of interaction kernels, and for several types of breakage
models.  We construct initial value and scaling solutions of the
fragmentation equations, and apply the "non-vanishing mass flux "criterion
for the occurrence of shattering transitions. These properties enable us
to determine the phase diagram for the occurrence of shattering states and
of scaling states in the phase space of model parameters.
\end{abstract}
\submitto{\JPA}

Fragmentation is a phenomenon of breakup of particles into a range of
smaller size particles. It is naturally found in a wide variety of
physical systems, ranging from comminution, breakup of grains, bubbles,
droplets, polymer degradation, disintegration of atomic nuclei, etc.
Fragmentation may occur through external forces, spontaneously, or
through interactions/collisions between particles. The subject has been
widely studied~\cite{Filippov}-\cite{collect-math}.

We are mainly interested in collision-induced nonlinear fragmentation as
caused by binary interactions. Such systems can be described by the time
evolution of $c(x,t)$, which is the number of particles of mass or size
$x$  at a given time $t$, or alternatively by its moments
$M_n(t)=\int_0^{\infty}dx x^n c(x,t)$. Quantities with similar properties
appear in coagulation processes. In either case the total mass is
conserved, $M(t)= M_1(t)=1$, while the total number of particles, $N(t)=
M_0(t)= \int dx c(x,t)$ is not. In irreversible coagulation, the mean
particle mass, $s(t)=M/N(t)$, increases monotonically, and may lead to a
finite time singularity at $t=t_c$, the gelation transition, characterized
by the appearance of an infinite cluster containing a finite fraction of
the total mass, $\Delta(t)=1-M(t)$ (order parameter), where $\Delta(t)
\neq 0$ for $t > t_c$. Alternatively, the gelation transition is
characterized by a non-vanishing mass flux $\dot{\Delta}(t) = -\dot{M}(t)$
from finite size particles (sol) to the infinite cluster
(gel)~\cite{Drake}-\cite{vDE}, i.e. a violation of mass conservation.

 In irreversible fragmentation the reversed scenario occurs. Here
$s(t)$ is monotonically decreasing, while the overall mass is conserved.
In these systems a finite time singularity may occur at $t_c$, the {\it
shattering} transition. It is characterized by a non-vanishing mass flux,
$\dot{\Delta}(t)$, i.e. the rate at which massive particles  are converted
into mass-less infinitesimals or fractal
dust~\cite{Filippov,Ziff-PRL,Ziff-PRL1, Szamel}.   If $\dot{\Delta}(t_c)$ is {\it
finite} this   transition has the character of a continuous phase
transition, described by the order parameter $\Delta (t)=1-M(t)$, as in
gelation~\cite{vDE}. In case $\dot{\Delta}(t_c) = \infty$   {\it all} mass
instantly 'evaporates' from the system; the transition is called explosive
and also referred to as a first order transition~\cite{KBN}.

  Smoluchowski's coagulation - fragmentation equation \cite{Drake} gives
the basic mean field description for reversible and irreversible
coagulation \cite{Ziff-PRL,Ziff-PRL1,vDE} and fragmentation processes
\cite{Filippov}-\cite{collect-math} in terms of the time evolution of $c(x,t)$ in spatially uniform (well
stirred) systems. In irreversible fragmentation or coagulation  the system
is described by a nonlinear coagulation rate, in combination with a
spontaneous linear fragmentation rate and/or    a collision- or
reaction-induced   nonlinear fragmentation rate. The system does not reach
a steady state, but at asymptotically large times  the distribution
function $c(x,t)$ approaches under rather general conditions to the
standard scaling form, which describes the typical $x-$dependence around
the mean particle size $s(t)$, which is steadily decreasing.

  The occurrence of  shattering has been addressed only partially in
the case of collision-induced nonlinear fragmentation. It shows a
behavior, qualitatively different from spontaneous (linear) fragmentation.
  Furthermore, the special cases analyzed so far are not necessarily
generic, but appear to be borderline cases.  In this letter we study the
occurrence of shattering for general classes of fragmentation models
within the framework of the nonlinear fragmentation equation and we
analyze its peculiarities and point out the parallels with gelation.

Collision-induced irreversible fragmentation can be described at the mean
field level by the nonlinear fragmentation equation with a collision term
$I$ composed of a loss and  a gain term\cite{Ch+Redner},
\ba \label{NL-frag-eq}
&&\partial c(x,t)/\partial t= I(x|c) \equiv
-c(x)\textstyle{\int^\infty_0}dy K(x,y) c(y)
\nn &&
+\textstyle{\int_x^{\infty}}dy \int_0^{\infty} dz b(x|y)K(y,z)c(y) c(z).
\ea
Here $b(x|y)$ is a {\it conditional} probability, describing the
distribution of outgoing fragments of mass $x$, given that a particle of
mass $y$ breaks \cite{Filippov,Ziff-PRL,Ziff-PRL1,Ch+Redner}.   One distinguishes:
(i){\it deterministic} or {\it splitting} models \cite{Ch+Redner,KBN},
where a particle breaks into two  equal fragments, hence $b(x|y) =2\delta
(x-y/2)$, and (ii){\it stochastic} models, where a fragment of  random
mass $x$ breaks off from a particle of mass $y$. As mass is  conserved in
a single breakup event, the outgoing fragment distribution has to obey
  the homogeneity requirement, $ b(x|y) = y^{-1}b(\frac{x}{y})$. For
simplicity, we take the standard form $b(s) = (\beta +2)s^\beta$
\cite{footnote1}, obeying,
\be \label{b-norm}
\textstyle{\int^y_0}dx x b(x|y)=y \quad,\quad \bar{N}= \int_0^y dx
b(x|y)=\textstyle{\frac{\beta+2}{\beta+1}}.
\ee
For physical reasons   the mean number of outgoing fragments satisfies
$\bar{N}\geq 2$  which implies $-1 <\beta \leq 0$. Binary breakup
corresponds to $\beta=0$.  In Eq.\Ref{NL-frag-eq} we consider binary
interactions, where the kernel $K(x,y)$ describes the interaction rate of
pairs of particles $(x,y)$. It may further contain a factor
$\bar{p}(x,y)$,   which gives   the probability that breakage indeed
occurs,   and   may depend on the masses $(x,y)$. If $\bar{p}$ is
constant, it can be absorbed in the time scale. Of importance in our
analysis is also the rate equation for the cumulative mass, $\dot{M}(x,t)
\equiv \int^\infty_x dy y \dot{c}(y,t)$,   as can be derived from Eq(1).
It reads,
\be \label{M-flux}
\dot{M}(x,t) = -\textstyle{\int_x^{\infty}}dy \textstyle{\int_0^{\infty}dz
\: y B(\frac{x}{y})} K(y,z)c(y)c(z)
\ee
  with   $B(s)=\int_0^s du u b(u)= s^{\beta +2}$. In applications
of the nonlinear Smoluchowsky equation a variety of collision kernels $K$
has been proposed  for processes induced by interacting particles
\cite{Drake,vDE}. Because of mathematical simplicity, kernels of the {\it
sum-product} form: $K(x,y)=1, x^p+y^p, (xy)^p, x^p y^q + x^q y^p$, etc
have been extensively studied in coagulation processes. Physically
motivated kernels are e.g. $K = (x^{-\alpha} +y^{-\alpha}) (x^{\beta}
+y^{\beta})$ with $\alpha=\beta=1/d$ for interaction rates among diffusing
particles, or $K\sim (R_x+R_y)^{d-1}$ for ballistic collision rates in a
$d$-dimensional system.  Here $R_x \sim x^{1/d}$ is the radius of a
particle of mass $x$, and $K$ a geometrical cross-section. In most cases
of physical interest, especially at limiting particle masses ($x\ll s(t)$
or $x\gg s(t)$), the kernels are continuous and homogeneous, i.e.
$K(ax,ay) = a^{\lambda} K(x,y)=a^{\lambda} K(x,y)$, and
$b(ax|ay)=a^{\lambda'}b(x|y)$ with $\lambda'=-1$. Kernels for coagulation
can be classified  by two exponents~\cite{vDE}, i.e. $K(x,y)\sim
x^{p}y^{q}$ if $x\ll y$, where $p\leq q$ and $\lambda=p+q$, with $p>0$
(class I), $p=0$ (class II) and $p<0$ (class III), with the physical
restrictions  $\lambda \leq 2,\: q\leq 1$~\cite{vDE}. This $(p,q)$-
classification  appears to be relevant for nonlinear fragmentation as
well, as we will show.

 The breakage probability, $\bar{p}(x,y)$, defines three different types
of models  depending on whether particle $x$ or $y$ breaks: (i) {\it
symmetric breakage}, where a randomly chosen particle of the interacting
pair $(x,y)$ breaks \cite{Ch+Redner,KBN}, and where $\bar{p}(x,y)=1$;
(ii){\it L-breakage}, where the larger particle breaks, hence
$\bar{p}(x,y)=\theta (x-y)$; (iii) {\it S-breakage}, where the smaller
particle breaks  and $\bar{p}(x,y)=\theta (y-x)$; $ \theta (x)$ stands for
the unit step function. The  corresponding nonlinear fragmentation
equation for L-breakage is obtained from Eq.\Ref{NL-frag-eq} by replacing
$I(x|c)$ with $I_L(x|c)$, with $K(x,y)$ replaced by $K_L(x,y) = K(x,y)
\theta(x-y)$, and similarly for S-breakage. Subsequently, we will discuss
the nonlinear stochastic fragmentation equation for kernels of class I, II
and III for symmetric, L- and S- breakage models. Regarding exact
solutions of the nonlinear fragmentation equation very little is known,
and mostly restricted to monodisperse initial conditions. The essential
references are \cite{Ch+Redner,KBN}, where the former contains a
representative list of the older literature. Ref.\cite{Ch+Redner} analyzes
the deterministic L- and S-breakage model, $ K_L(x,y)= x^p \theta(x-y)$
and $K_L(x,y)= x^p \theta(y-x)$, and Ref.\cite{KBN} does so for both the
deterministic and the stochastic breakage models with $ K_L(x,y)=
\theta(x-y)$ and $K_L(x,y)= \theta(y-x)$.

Regarding the structure of the nonlinear integral-differential equations
(1) and (3) for cases where $K(x,y)=a(x)a(y)$ is a  general product
kernel, it has been observed \cite{Ch+Redner}, that the nonlinear
fragmentation equation can be transformed into a linear one with a new
time variable $\tau (t)$ that is related to the physical time $t$ in a
nonlinear manner. The functional form of $\tau(t)$ determines whether a
shattering transition is present or absent. So, to explain this
dependence it is paramount to
discuss how the initial solutions $c(x,t)$ of the nonlinear fragmentation
equation for a given $c(x,0)$ can be constructed from the initial
solutions $\bar{c}(x,\tau)$ of the linear fragmentation equation. To this
end, we analyze {\it linear} fragmentation,
\be \label{lin-frag-eq}
\partial c(x,t)/\partial t=-a(x)c(x)+\textstyle{\int_x^{\infty}}
dy b(x|y)a(y)c(y),
\ee
where $a(x)c(x)$ represents the spontaneous or externally-induced linear
breakup rate.  Exact solutions $c(x,t)$ are known for algebraic
fragmentation rates, $a(x)=x^\alpha$ for all real $ \alpha$, and
mono-disperse initial conditions, $c(x,0)=\delta(x-x_0)$
\cite{Filippov,Ziff-PRL,Ziff-PRL1,Williams}. These $c(x,t)$'s are the causal Green
functions of Eq.\Ref{lin-frag-eq} with a monomer source $\delta (t)\delta
(x-x_0)$~\cite{Williams}.  So, $c(x,t)=0$ for all $x > x_0$ at $t>0$. In
the sequel we set $x_0=1$. Spontaneously fragmenting systems
\cite{footnote2} with $\alpha \geq 0 $ are non-shattering, i.e. the total
mass $M(t)=1$ at all times, and the total number of particles, $M_0(t) <
\infty$ for all $t<\infty$. Moreover, moments with $n+\beta+1>0$ exist,
and evolve for large $t$ as $M_n(t) \sim t^{(1-n)/ \alpha}$, while those
with $n+\beta+1\leq 0$ are divergent. On the other hand, spontaneously
fragmenting systems with $\alpha<0$, are shattering, and mass loss starts
at the initial time. So, shattering occurs at $t=t_c=0$, where $
\dot{\Delta}(t_c) $ is  finite; hence the transition is continuous. The
possibility of an explosive shattering transition with $\dot{\Delta}(t_c)
= \infty$ is never realized. All initial solutions, which have by
definition $t>0$, are non-mass conserving post-shattering solutions with
$t>t_c=0$. They behave for small $x$ as $c(x,t) \sim A(t) x^{-\theta}$
with $\theta = \alpha+2$. Consequently, moments $M_n(t)$ with $n\leq
1+\alpha$ are divergent for {\it all} times, and those with $n > 1+\alpha$
decay for long times as $M_n(t) \propto A(t) \propto e^{-t}
t^{(\beta+2)/\alpha}$, including $M(t)$. We also point out that for
$\alpha>0$ the exact solutions converge asymptotically to {\bf a}  {\it
standard} scaling form, $c(x,t) = (1/s^2(t))\varphi
(x/s(t))$\cite{Filippov}-\cite{collect-frag}, where the scaling limit is
formally defined as the coupled limit, $t \to \infty$ and $x \to 0$ with
$x/s(t)$ kept constant. In this limit, where $s \sim 1/M_0 \sim
t^{-1/\alpha} \to 0$,(i.e. the total number of particles $M_0(t)$
diverges) the exact solution becomes, $s^2 c(s u,t) \sim \varphi (u) \sim
u^\beta\exp[-u^\alpha ]$. Those with $\alpha < 0$ do not approach a
scaling form. Inspection of the exact solution as $\alpha \to 0$
at fixed $(x,t)$ shows that $c(x,t)=0$ for all $x>x_0=1$, and reads for
$x<1$ (see Ref.~\cite{Ziff-PRL}), 
\be \label{sol-except}
c(x,t)= e^{-t}( 2t/\ln(1/x))^{1/2} I_1\left[2(2t\ln(1/x))^{1/2}\right]
\ee
where $s(t)=e^{-t}$, and $I_1(x)$ is the modified Bessel function of
integer order $n=1$ . This expression shows that the borderline case, $
\alpha =0$, is exceptional, i.e. non-scaling {\it and} non-shattering.

 Let us now consider the {\it nonlinear} fragmentation equation for
symmetric breakage with product kernel $K(x,y)=a(x)a(y)=(xy)^p$ and $0\leq
\lambda=2p\leq 2$. In this case Eq.\Ref{NL-frag-eq} is a quasi-linear
equation for which exact initial value solutions can be obtained. It 
can
be mapped onto the linear fragmentation equation with $\alpha \to p $ and
$t \to \tau $, defined through $d\tau = M_p(t)dt$. Consequently the mass
distribution, $\bar{c}(x,\tau)$ and moments $\bar{M}_n(\tau) = \int^1_0 dx
x^n
\bar{c}(x,\tau)$ for mono-disperse initial values are known explicitly,
and only $\tau(t)$ needs to be determined in order to have the complete
solution as a function of $t$. For $p=\alpha>0$, where $K$ is a class I
kernel, total mass is conserved for all $\tau$, and  the moments for
$n\neq 1$ behave at large $\tau$ as $\bar{M}_n(\tau)\sim \tau^{(1-n)/p}$.
Furthermore, to have $\tau$ as a function of $t$ we need to invert the
relation $t=\int^{\tau}_0 ds/\bar{M}_p(s)$. If $\lambda=2p>1$, then
$t(\tau)\sim \tau^{2-1/p}$ is monotonically increasing, the relation is
invertible, and $t\rightarrow \infty$ as $\tau\rightarrow \infty$. Hence,
$M_1(t)=1$ for all $t$, and there is no shattering and no divergence of
$M_0(t)$ at any finite time. However, if $\lambda=2p<1$, then as $\tau \to
\infty$, $t \to t_c \equiv\int_0^{\infty} d\tau/M_p(\tau)<\infty$.
Consequently there exists a finite time singularity,
$\tau\sim(t_c-t)^{-p/(1-2p)}$ as $t\to t_c$ and mass remains conserved
only for $t<t_c$ and vanishes instantaneously at $t_c$, where
$\dot{\Delta}(t_c) = - \infty$. At the same time all moments, behaving as
$M_n(t) \sim (t_c-t)^{(1-n)/(1-2p)}$, either diverge or vanish. These are
the hallmarks of an explosive shattering transition at $t_c$ where all
massive particles are converted instantaneously into fractal dust. For $
\lambda  =2p <0$ (class III)  the kernel $K=(xy)^p$  can be mapped on the
linear equation through $d \tau = M_{-|p|}(t) dt$. Its moments
$\bar{M}_n(\tau)$ with $n <1+\alpha= 1-|p|$ do not exist. Consequently
$\tau(t)$ is not defined, and $c(x,t)$ does not exist for mono-disperse
initial conditions with class-III kernels. The same applies to scaling
solutions. The corresponding class II kernel with $p= \alpha = 0$   or
$K=1$ represents an exceptional  point, as discussed below Eq.(5). The
solution at $p=0$ ($K=1$, class II) exists, is shattering, and
$\bar{c}(x,\tau)$ is identical to the non-generic, non-scaling solution of
the linear Eq.\Ref{lin-frag-eq} at $\alpha=0$.

 To determine possible {\it scaling} solutions we substitute the scaling ansatz
$c(x,t) = (1/s^2)\varphi (u = x/s)$ in  \Ref{M-flux}, and take the
derivative, yielding  the  scaling equation for symmetric, L- and
S-breakage $ ( A=0,L,S)$,
\be \label{standard-scaling}
\left({\varphi(u)}/{u^{\beta}} \right)^{\:\prime} = -(\varphi(u)/\gamma
u^{\beta +1})\textstyle{ \int^\infty_0 }dv K_A(u,v)\varphi (v),
\ee
where $\gamma$ is an arbitrary positive separation constant,  $\varphi(u)$
has to satisfy the boundary condition, $u^2
 \varphi(u) = 0$ as $u \to \infty$, and $K_0=K$. For all three types of breakage
models the evolution equation for the mean particle size is the same,
$\dot{s}s^{-\lambda} = -\gamma$. Its solution is,
\be \label{s-t}
s(t) \sim \left\{ \begin{array}{ll}
 (t_0+t)^{1/(1-\lambda)}   & \mbox{for} \lambda>1, t\to \infty \\
\exp[-\gamma t]      &  \mbox{for} \lambda=1, t\to \infty \\
 (t_c-t)^{1/(1-\lambda)} &  \mbox{for} \lambda< 1, t \leq t_c
\end{array} \right.
\ee
where $t_0, t_c \to \infty $ as $ \lambda \to 0$. The appearance of the
finite time singularity at $t_c$ indicates that shattering only occurs
for $\lambda <1$. Systems with $\lambda \geq 1$ are non-shattering
\cite{Ch+Redner}. Note that the scaling limit in the pre-shattering
critical region is defined as the coupled limit: $t \uparrow t_c$ and $x
\to 0$ with $x/s(t)=$  constant .

For sum-product kernels in symmetric breakage the rhs in
Eq.(\ref{standard-scaling})   reduces to sums of powers $u^s$, multiplied
by coefficients $m_n = \int^\infty_0 du u^n \varphi (u)$, which can be
determined self-consistently. Specifically, for $K(x,y)=(x y)^p$ ($p>0$,
class I), one obtains from the rhs of (6) $\psi^\prime (u) = \psi (u) m_p
u^{p-1}/\gamma$, where $\psi(u) \equiv \varphi(u)/ u^\beta$. This can be
readily integrated to yield,
\be \label{pp-sol}
\varphi(u) = C u^\beta \exp[ - u^p m_p/\gamma p],
\ee
where $C$ and $\gamma$ are determined by imposing  normalization ($m_0=1$)
and  mass conservation ($m_1=1$). In order to get simpler analytic
expressions, we use the invariance property that the scaled distribution,
$\bar{\varphi}(\bar{u})$, obtained under the similarity transformation
$\bar{\varphi}(\bar{u}) =s_0^{-2} \varphi(u /s_0)$ for an arbitrary
constant $s_0$ also satisfies Eq.\Ref{standard-scaling} with $\bar{m}_1
=1$.  This property allows us to fix  $\gamma$ by setting $m_p/\gamma p
=1$, which is a self-consistency equation. With this choice the moments
read $m_n = (C/p)\Gamma (b_n) = \Gamma(b_n)/\Gamma(b_1)$ where $\Gamma(x)$
is the Gamma function and $b_n= (1+\beta+n)/p$. Hence, the scaling
distribution function can be expressed as $\varphi (u) =  p u^\beta
e^{-u^p}/\Gamma ((\beta+2)/p)$. Solutions with a different normalization,
e.g. $m_0=1$, are easily derived using the invariance property under
similarity transformations.

Similarly one derives that the size distribution for sum-kernels,
$K=x^p+y^p$ (class II) with $p>0$, has the form $\varphi(u)= C
u^{\bar{\beta}}e^{-u^p} =p u^{(\beta-1)/2)}e^{-u^p} /
\Gamma((\beta+3)/2p)$ with $\bar{\beta} =
\beta - m_p/\gamma$. Here we impose the self consistent equation
$m_0/p\gamma =1$ to determine the separation constant $\gamma$. Imposing
mass conservation leads to the second equality in the previous equation.
The moments of the distribution can then be computed; in particular $m_p =
b_0\gamma p$ where now $b_n = (1+\bar{\beta} +n)/p$ which implies
$\bar{\beta} = \half(\beta -1)$. The exact scaling solutions for the
symmetric breakage kernels $(p,p)$ and $(0,p)$ above  approach different
limiting forms as $p \to 0$. So, the analysis starting below Eq.
\Ref{sol-except} shows that the $K-$kernel with $(p,q)=(0,0)$ is quite
singular. In a similar manner the scaling solutions $\varphi (u)$ for the
geometric collision cross-section, $K\sim (x^{1/3}+y^{1/3})^2$, and
closely related kernels can also be found, as well as the asymptotics of
$\varphi(u)$ for general class I and II kernels in all breakage models of
type $A =(0,L,S)$. In L- and S-breakage models for {\it generic} $K$ no
exact initial value or scaling solutions are known, except for the
non-generic borderline case $K=1$ in Ref.\cite{KBN}, which lacks standard
scaling in the variable $u=x/s(t)$ in all three breakage models.

 To analyze from a broader perspective the occurrence of shattering, we
will focus on the behavior of the cumulative mass flux for vanishingly
small masses. If $\lim_{x \to 0} \dot{M}(x,t) \equiv \dot{M}(t)$ is {\it
vanishing} at $t_c$, the system is non-shattering; otherwise there is
shattering. If $-\infty < \dot{M}(t_c)< 0$, the phase transition is {\it
continuous}, and $c(x,t)$ exists for $t>t_c$. If $\dot{M}(t_c)= - \infty$,
the phase transition is explosive (first order), and $c(x,t)$ does not
exist for $t>t_c$.

 For general $K(x,y)$  in class I, II and III   the mass flux
$\dot{M}(x,t)$  can only have a non-vanishing limit for $x \to 0$  if
$c(x,t)$ is of power law type, because the rhs contains the factor
$x^{\beta+2}$ with $-1<\beta \leq 0$. So we propose the post-shattering
ansatz $c(x,t) \sim A(t) x^{-\theta} \:(x \to 0)$, and determine $\theta$
such that $\dot{M}(t) \neq 0$ (see Refs.\cite{ZHE,vDE}). Moreover $\theta
<2 $ because the total mass  should remain finite.

The evolution equation for fragmentation with S-breakage is described by
Eq.(\ref{M-flux}) with $K$  replaced by $K_S$. Inserting the ansatz above
yields for small $x$,
\ba \label{M-flux-S}
&&\dot{M}(x,t)\simeq -A^2 k(\theta)
(x^{3+\lambda-2\theta})/({2\theta+\beta-1-\lambda})
\nn && k(\theta)= \textstyle{\int^\infty_1 ds K(1,s) s^{-\theta}
\qquad ( 1+p <\theta <2) }
\ea
where $K(1,s) \sim s^p$ for $s \gg 1$.  In case $\theta
=\half(3+\lambda)$ the above small-$x$ limit yields a finite result for
the mass flux $\dot{M}(t)=-{A^2(t)k(\theta)}/(2+\beta)$, i.e. it allows
the existence of a {\it continuous} shattering transition with a
post-shattering solution of algebraic form for  $t>t_c$; $\theta<2$
implies $\lambda<1$. At the (unknown) shattering time $t_c$ mass
conservation breaks down, and for $t>t_c$ there exists a non-vanishing
order parameter $\Delta(t)= 1-M(t)>0$ with $\dot{\Delta}(t)\sim A^2(t)$.
Eq. (\ref{M-flux-S}) includes also the special result, obtained in
\cite{KBN} for the S-breakage model with $K=1$ and $\beta =0$.

It is remarkable that the post-gelation distribution, $c(x,t)\sim
A(t)x^{-\theta}$, occurring in Smoluchowski's coagulation equation for
$\lambda>1$, has the same exponent $\theta = (3+\lambda)
/2$~\cite{vDE,ZHE} as in the fragmentation process above; hence the close
analogy between gelation and continuous shattering. Note that the value
of exponent $\beta$ has no influence on the existence of shattering.

A similar analysis can be performed for symmetric and L-breakage models.
In doing so we introduce a lower cut off $\epsilon y$ on the $z$-integral
in Eq.~\Ref{M-flux}, and take $\lim_{\epsilon \to 0}$ at the end of the
calculations. Due to the physical restrictions on the allowed values for
$p$ and $q$, shattering is always explosive rather than continuous.

From the properties discussed in this letter we can construct the phase
diagram for symmetric breakage in the $(p,q)-$plane.  It is restricted to
the triangular region, spanned by $ (0,0), (0,1), (1,0)$ and includes the
boundaries. The region with $ 0 \leq \lambda <1$ represent shattering
systems, and the region with $ 1 \leq \lambda \leq 2$ represents
non-shattering ones. The whole triangular region shows standard scaling in
the variable $u = x/s(t)$, except in the singular corner $(0,0)$.
Regarding the phase diagram for L- or S-breakage the location of the left
boundary (separatrix between "non-existence" and "existence of scaling
solutions"), including the singular point $(0,0)$, is unknown, and the
behavior on it may be different from its right and left limit. From
Ref.\cite{KBN} it is known that a new type of scaling in the variable
$x/m^*(t)$ appears at $(0,0)$. Here $m^*$ is a characteristic mass, that
cannot be defined a priori, but follows from a clever mapping of the
fragmentation equation on the nonlinear equation for travelling fronts.
Contrary to models with symmetric breakage, which are quasi-linear, the
scaling equations for L- and S-breakage are genuinely nonlinear, i.e.
$\varphi^{\prime\prime}(u) = F(u,\varphi^\prime, \varphi)$, and the only
solutions known are those for the singular point $(0,0)$.

We have discussed the generic behavior of  collision-induced irreversible
fragmenting systems at the mean-field level. We have shown that the
scenarios for nonlinear fragmentation are qualitatively different from
those of spontaneous linear fragmentation. The behavior of the shattering
transition depends both on the kind of fragmentation kernel and on the
type of breakage. For symmetric and L-breakage, where the kernel $K$ has a
degree of homogeneity $\lambda <1$, shattering is always explosive, while
S-breakage models show a continuous shattering transition, analogous to
gelation. The existence of a transition does not depend on the details of
the fragment distribution, $b(x|y)$, i.e. on $\beta$. Shattering  in
collision-induced fragmentation always takes place at a finite time
$t_c\neq 0$, as opposed to linear fragmentation where shattering occurs at
$t_c=0$ for $\alpha<0$. Contrary to gelation~\cite{vDE}, in class III
kernels with symmetric breakage neither initial, nor scaling solutions exist. The
solutions for fragmentation models with a fragment distribution, $b(s)=0 $
for $s<s_0$, $b(s) \neq 0$ for $s_0<s<1$, have scaling solutions
$\varphi(u) $ exhibiting a log-normal distributions at small
$u$~\cite{Ch+Redner}.

\ack M.E. acknowledges the PIV program from DURSI (Generalitat de
Catalunya) for financial support. I.P. thanks DGICYT of the  Spanish
Government and Distinci\'o from DURSI (Generalitat de Catalunya) for
financial support.


\begin{thebibliography}{99}
\bibitem{footnote2} The analytic results in this paragraph have been
calculated with the help of Ref.\cite{Szamel}
\end{thebibliography}
\end{document}